\documentstyle[preprint,eqsecnum,aps]{revtex}
\tighten
\begin{document}
\draft

\title{  Vortices and domain walls in a    Chern-Simons    theory
with magnetic moment interaction. }
\author{  Armando Antill\'on $^a$\footnote{e-mail: armando@ce.ifisicam.unam.mx},
 Joaqu\'in Escalona$^b$\footnote{e-mail: joaquin@servm.fc.uaem.mx},
 \\and  
Manuel Torres$^c$\footnote{e-mail: manuel@teorica1.ifisicacu.unam.mx  }}
\address{ $^a$ Laboratorio de Cuernavaca, Instituto de F\'isica,  
Universidad Nacional  Aut\'onoma \\
de M\'exico,   Apdo. Postal 48-3,  
62251 Cuernavaca, Morelos,   M\'exico. \\
$^b$ Facultad de Ciencias, Universidad Aut\'onoma del Estado de Morelos, \\
Apdo. Postal 396-3, 62250 Cuernavaca, Morelos,   M\'exico.\\
 $^c$Instituto de F\'isica,  Universidad Nacional Aut\'onoma  de M\'exico, \\
Apdo. Postal 20-364, 01000  M\'exico, D.F., M\'exico.}

\maketitle
\begin{abstract}
We study  the structure and properties of vortices in a recently 
proposed  Abelian  Maxwell-Chern-Simons model  in $2 +1 $ dimensions.
The  model  which is described by gauge field interacting with a complex scalar field, includes two parity and time violating terms: the Chern-Simons  and  the 
anomalous magnetic terms. Self-dual relativistic  vortices are discussed in detail.
 We also find one dimensional soliton solutions of the domain wall type. The vortices are  correctly described by the  domain wall solutions in the large flux limit. 
\end{abstract}

\narrowtext

\section{Introduction}\label{introduction}\indent

Of the gauge field  theories, the self-dual theories deserve
special attention.  Self-duality refers to theories in which 
the interactions have particular forms and special strengths such 
that the equations of motion reduce from  second- to first-order  differential equations; these configurations  minimize a  functional, often  the energy \cite{review}. 
For example the  Abelian-Higgs model  admits topological 
solitons of the vortex type \cite{nielsen}. In this model  the scalar potential is of the form $V(\phi)  \sim (|\phi|^2 - v^2)^2$ and the vortices satisfy a set of Bogomol'nyi or self-dual equations when   the vector and scalar masses
 are chosen to be equal  \cite{bogo,devega}.  The self- dual point corresponds to the boundary  between type-I and type-II superconductors.  In this  point  the vortices  become  non-interacting and static multisoliton solutions may  be expected \cite{jacobs,taubes}. 
We also notice  that the   self-dual structure of the  gauge theories 
 is related at a fundamental level  to the existence of an  
extended supersymmetry \cite{super}.

Recently   considerable interest has been paid to  the study of vortex solutions 
in  $(2 + 1)$ dimensional Chern-Simons (CS) gauge theories.  
One common feature of the Chern-Simons solitons is that they carry 
electric charge as well as   magnetic flux   \cite{paul1}, in contrast with the  electrically neutral  Nielsen-Olesen vortices.  In addition they possess
fractional spin; a  property that  is fundamental  to  consider them  as candidates  for anyon like objects in quasiplanar systems.
Self-dual Chern-Simons theories are known to exist if one considers   pure CS theories.  In these theories   the Maxwell term in 
the Lagrangian  is absent and   the dynamics  for the gauge field is solely provided by the Chern-Simons term  \cite{deser,review2}.
The self-dual Chern-Simons theories enable a realization with either relativistic
\cite{solnore}  or nonrelativistic \cite{hoja,jack2} dynamics for the matter degrees of freedom.
In the case of  the relativistic  theory  self-dual topological and nontopological vortex solutions have been found with a particular  sixth-order potential  of the form 
  $V(\phi)  \sim  |\phi|^2 (|\phi|^2 - v^2)^2$  when the vector and scalar masses
are equal \cite{hoja,jack2}.

The question can be posed as to  whether there are   self-dual 
models  in which the gauge field Lagrangian includes 
both the Maxwell and the Chern-Simons term.   A self-dual  Maxwell-Chern-Simons  gauge theory can be constructed  if  a magnetic moment interaction 
is added
between the  scalar  and the gauge fields \cite{torres}.
\footnote{The author of  reference \cite{leelee} analyzed the Abelian Chern-Simons model with  a sixth-order potential when the Maxwell term is included. In this case it is necessary to  add a neutral scalar field to obtain the self-dual condition}
 If the  interest is pursuit  in a low energy 
effective theory containing at most second-order derivative terms,  such a magnetic moment  interaction has to be included.    Two steps are followed  to obtain the self-dual limit.
First a particular  relation between the   CS mass and 
the anomalous magnetic coupling  is stablished whereby 
the equations for the gauge fields 
reduce from second- to first-order differential equations 
similar to those of the pure CS theory. 
Second,  if the scalar  potential is selected as  a simple $\phi^2$ potential 
and the scalar mass is made equal to  the   topological mass, the energy obeys a Bogomol'nyi-type lower  bound,  which is saturated by fields satisfying self-duality equations.  The potential possesses a unique minimum at $\phi = 0$ and topological solitons certainly do not exist, yet the theory allows nontopological vortex configurations. 
In this paper we  examine the  theory and the properties of these nontopological vortices in more detail. In addition we  find that the model admits one-dimensional soliton  solutions of the domain wall type.  The domain wall carries both 
magnetic flux and electric charge  per unit length. Furthermore  we find  that the domain wall configurations provide an approximate solution to the self-dual vortices in the large flux limit.   It must be emphasized that the existence of  domain wall solutions is unexpected, because the scalar potential  has a unique nondegenerate minimum. 

As mentioned above there are several aspects of the $\phi^2$ Maxwell-Chern-Simons  gauge theory which justify  further consideration.  In section  \ref{model} we introduce the model  in which  a charged scalar field  is coupled via a generalized covariant derivative  to the gauge field whose dynamics includes  both the Maxwell and the Chern-Simons term. Initially we consider an arbitrary renormalizable scalar potential in $(2 +1 )$ dimensions in order to discuss properties of the theory both in the symmetric and  in the spontaneously broken phase.  As  first pointed out in reference \cite{paul2}, the nonminimal term in the covariant derivative combined with the spontaneous
symmetry mechanism induce a kind of Chern-Simons term. 
However we demonstrate that the  induced CS term behaves in the same way as the explicit CS term only in the topological trivial sector of the theory. 
The properties of the two terms are different in the topological non-trivial 
sector, in particular the induced CS term  does not contribute  to the fundamental 
relation between charge and magnetic flux.   Rather, the  magnetic moment induces
a contribution to the magnetization of the vortex that is proportional to the charge of the configuration. In section  \ref{smass}  we study the propagating 
modes for the vector field, which consist of two longielliptic waves with different values for the masses.   Then in section \ref{pure}  we discuss
the conditions required  to reduce the original  gauge field equations to equations of the pure Chern-Simons type.
Section \ref{nontopo} is devoted to the derivation of the self-duality equations, 
and to the detailed analytical and numerical
study of the cylindrically symmetric vortex solutions. 
In section \ref{domwa} we discuss domain wall solutions with finite energy per unit length and use these to further examine  the properties of the vortex solutions in the large flux limit. 
Concluding remarks comprise the final section.

\section{The model }\label{model}\indent

Our  model  possesses  a local $U(1)$ symmetry and is described by the following effective Lagrangian 

\begin{equation}\label{lag}
 {\cal L} = -{1 \over 4} F_{\mu \nu} F^{\mu \nu} \, + 
   {\kappa \over 4} \epsilon ^{\mu \nu \alpha} A_\mu F_{\nu \alpha}  
\,    + {1 \over 2} |{\cal D}_\mu \phi|^2 - V( |\phi | ) \, ,
\end{equation}
where  $\kappa$ is the topological mass, 
$F_{\mu\nu} = \partial _{\mu} A_{\nu} - \partial _{\nu}A_{\mu}$. 
We use natural units $\hbar = c = 1$ and the Minkowski-space metric is 
 $g_{\mu\nu} = diag(1,-1,-1); \, \mu=0,1,2$. The covariant derivative 
is generalized as 
 
\begin{equation}\label{dercov}
{\cal D}_{\mu} = \partial _{\mu} - {ieA_\mu}
- i{g\over 4} \epsilon_{\mu\nu\alpha}F^{\nu\alpha} 
\equiv \partial _{\mu} - {ieA_\mu}
- i{g\over 2} F_{\mu}\, , 
\end{equation}
where  we have defined the dual field 

\begin{equation}\label{fmu}
F_\mu \equiv {1 \over 2} \epsilon_{\mu\alpha\beta}   F^{\alpha\beta}
\, .  \end{equation}
In terms of the dual field the CS term  takes the simpler
form $(\kappa /2) A_\mu F^\mu$.
The introduction of an  anomalous magnetic  term in the covariant derivative is consistent with   the Lorentz and the gauge invariance of the theory; however   it breaks the ${\cal P}$ and
${\cal T}$ symmetries.  An specific feature of a $(2 +1 )$ dimensional world is  that a  Pauli-type coupling  ($i.e.$ a magnetic coupling) can be incorporated into the covariant  derivative, even for spinless  particles \cite{paul2,stern,torres}.
In fact,  in  reference \cite{kogan} it was demonstrated that radiative corrections can induce a magnetic coupling for anyons, proportional to the fractional spin.  The  electromagnetic interactions of  charged anyons, in particular its magnetic moment, have also been discussed for point-particles in   $(2 +1)$ dimensions  using the appropriate representations of the Poincar\'e group \cite{chou}. 
 
The most  general renormalizable potential in $(2+1)$ dimensions is of the form 

\begin{equation}\label{pot}
V(\phi)  = a_6 |\phi|^6 +  a_4 |\phi|^4 + a_2 |\phi|^2 \,  .
\end{equation}
As we shall see later, the particular second-order form of the scalar potential 
together with its overall strength are fixed by self-duality condition.
For the time being   we leave the parameter in (\ref{pot}) free in order to discuss
both the broken and the unbroken phases of the theory.

As  Paul and Khare \cite{paul2} point out a   CS  term can be generated by spontaneous symmetry breaking. However,  the properties of this CS term are not the same as those of the explicit CS term appearing in  Eq. (\ref{lag});  we would like to understand the origin of these differences.  
 Suppose the  potential  is selected to  have symmetry breaking  minimum at $|\phi| = v$. Then  in terms of the gauge invariant potential 

\begin{equation}\label{ainv}
\tilde{A}_\mu \, = \, A_\mu - {1 \over e} \partial_\mu Arg (\phi ) \, , 
\end{equation}
the contribution of the covariant derivative  to the Lagrangian evaluated in   the vacuum  configurations ($|\phi| = v$)  can be written as

\begin{equation}\label{dermin}
{1 \over 2} |{\cal D}_\mu \phi|^2 = {1\over 2}
\left[e^2v^2 \tilde{A}_{\mu} \tilde{A}^{\mu}
+{g^2 v^2\over 4} F_\mu F^\mu +
 egv^2  \tilde{A}_{\mu} F^\mu  \right]\, .
\end{equation}
The first term in this expression is the usual gauge field  mass ($M=ev$)  generated by the spontanous symmetry breaking.  The second term modifies the 
coefficient of the  Maxwell term in the Lagrangian. Finally,  the last term  is a kind of  CS type term generated by  the spontaneous symmetry breaking with topological mass  $ ev^2g$.   However    the explicit CS term  is   of the form
$(\kappa /2) A_\mu F^\mu$.  Instead,  in the   induced CS term $F^\mu$
 couples to  $\tilde{A}_\mu$ rather than   to  $A_\mu$. The gauge field $\tilde{A}_\mu$  is massive so it  has a finite correlation length. This does not imply that $A_\mu$ should also  fall off exponentially; it can  remain  a pure gauge.  Indeed  this is the  case  around   a vortex where the  long range contribution, which is  locally pure gauge, is globally non-trivial  giving rise  to a non-vanishing magnetic flux. 
One of the effects  of  the explicit CS term is that  a vortex with magnetic flux $\Phi$ must also carry electric charge $Q$, with the two quantities related as 
$Q = -  \kappa \Phi $. The induced  CS mass term $ ev^2g$ does not contribute to this relation because of  the finite correlation length of $\tilde{A}_\mu$. Consequently,  we conclude that  in  topologically non-trivial sector 
 the induced term  ${\tilde A}_\mu F^\mu$ in   (\ref{dermin}) 
has not the same properties as those of  the  CS term; so it cannot be considered
 a genuine CS term. Only   in the topologically  trivial sector does  the induced term
have the same properties as those of the explicit    CS term.

The equations of motion  for the Lagrangian in  (\ref{lag})   are 

\begin{equation}\label{eqm1}
{\cal D}_\mu {\cal D}^\mu \phi 
        =  -  2 {\delta V \over \delta \phi^*}   \,  , 
\end{equation}

\begin{equation}\label{eqm2}
 \epsilon_{\mu \nu \alpha}  \partial^\mu \big [ F^\alpha +
{ g \over 2 e} J^\alpha \big]   
     =  J_\nu - \kappa F_\nu   \,  ,
\end{equation}
where  the conserved Noether current is given as

\begin{equation}\label{cur1}
J_{\mu} = - {ie\over 2} 
             \left[\phi ^{*}({\cal D}_{\mu}\phi) -
               \phi ({\cal D}_{\mu}\phi)^{*} \right]\, .
\end{equation}

From the  equation of motion (\ref{eqm2})  it is clear that we can define a current ${\cal J}_{\mu}$  that is also conserved. ${\cal J}_{\mu}$  is  defined by

\begin{equation}\label{cur2}
{\cal J}_{\mu}= J_{\mu} 
+ {g\over 2e}\epsilon _{\mu\nu\alpha}\partial ^{\nu}J^{\alpha} \, . 
\end{equation}
If the    current $J_\mu$ is bounded or 
vanishes  faster than $ 1/r$  at spatial infinity, then the charges calculated from 
$J_0$ and ${\cal J}_0$ coincide \cite{nota}.  

 Let us  further examine the gauge field  equations of motion (\ref{eqm2})
expressed  in terms of  the electric and magnetic fields
 $E_i = F_{0i}$  and $B= F^{12}$ respectively;   they  read 

\begin{eqnarray}\label{gauss}
\nabla\cdot\vec E - \kappa B &=& \rho 
+ {g\over 2e}\epsilon _{ij} \partial ^i  J^j \, ,
\nonumber \\
\epsilon _{ij} \left( \partial ^j B + \kappa  E^j  \right) &=&
J_i + {g\over 2}\epsilon _{ij}\partial ^j\rho
+ \partial ^0E_i\, . \end{eqnarray}
These equations   can be identified as the modified Gauss and Ampere laws respectively.
One of the  most important consequences of the CS term is the fact that any object with  magnetic flux $ \Phi = \int d^2x B$   also carries electric charge  
$Q = \int d^2 x \rho $,  with the two quantities related as 

\begin{equation}\label{funda}
Q \, = \, - \kappa \Phi  \, . 
\end{equation}
Indeed    integrating the Gauss law (\ref{gauss}) over all   space we find that 
the contribution  of  $\nabla \cdot  \vec E$ is zero, because of the long-distance
damping produced by the   ``photon'' mass.  Similarly
the integral  of the last  term also vanishes; this is true even if  the  symmetry 
is  spontaneously broken. The reason, as explained earlier,  is that   in the 
Higgs vacuum the current  $J_\mu$ makes use of the  massive field 
${\tilde A}_\mu$ instead of $A_\mu$.   Hence,  we obtain the desired result in 
(\ref{funda}).

While the magnetic moment $g$ does  not  have a direct  effect on the fundamental relation (\ref{funda}), it  does produce interesting effects,  one of which  can be found in the magnetization
of the excitations of the system.
The magnetization  is found by  coupling an external magnetic field and extracting the linear coupling from  the   Lagrangian (\ref{lag}). It is given by

\begin{equation}\label{magne1}
M  \, = \, \int  d^2 x  \left( \epsilon_{ij} x^i J^j \right)  \, ,
\end{equation}
 where  $\vec J$ is the  matter current in (\ref{cur1}). Utilizing the modified Ampere law  in  (\ref{gauss})  we find   for  a static   configuration

\begin{equation}\label{magne2}
M = \Phi - {g \over 2 e } Q - {\kappa \over 2 }   \int d^2 x \left( \vec r \cdot \vec E  \right)
\, . \end{equation} 
In the absence of parity breaking terms the magnetization  is equal to the magnetic flux, a result known for the neutral Nielsen-Olesen vortices \cite{nielsen}. Here
$M$   has two extra pieces:  the magnetic moment    $g$ which  induces a contribution proportional to the charge of the configuration; and 
 the  term proportional to $\kappa$ which, 
 unlike  the two first contributions,  depends on the structure factor of the vortex configuration  so cannot be explicitly integrated.

Finally, the energy momentum tensor is obtained by varying the curved-space form 
of the action with respect  to the metric

\begin{eqnarray}\label{tmunu}
   T_{\mu \nu}    & = &    \left( 1 - {g^2 \over 4} |\phi|^2 \right)
 \left( F_\mu F_\nu  -  {1 \over 2} g_{\mu \nu} F_\alpha F^\alpha \right) 
     \nonumber\\
 \nonumber\\
    & +  & {1 \over 2}\left( \nabla_\mu \phi \left( \nabla_\nu \phi \right)^\ast 
- g_{\mu \nu} \left[{1 \over 2} |\nabla_\lambda \phi|^2 -  V(|\phi|) \right]
+ H. c. \right)  
   \,,   \end{eqnarray}
where $\nabla_\mu = \partial_\mu - i e A_\mu$  only includes the gauge
potential contribution.  The Chern-Simons  and linear terms in $g$ 
do not appear explicitly in $T_{\mu\nu}$. This is  consequence of  the fact that  these  terms  do not make use of the space-time  metric tensor $g_{\mu\nu}$, thus    when  $g_{\mu\nu}$ is varied  to produce $T_{\mu\nu}$  no contributions  arise  from  these terms \cite{review2}. 
The expression  (\ref{tmunu}) can be considered as  the energy momentum  tensors of an
Abelian  Higgs model in which the Maxwell term is multiplied by a particular dielectric function of the form  $\left( 1 - {g^2 \over 4} |\phi|^2 \right)$ \cite{gosh}.

\section{Propagating  modes }\label{smass}\indent

 The model described in the previous section has in general 
three propagating modes in the vacuum state,  one of which   corresponds to   the scalar field. 
In order to describe the particle content of the gauge field degrees 
of freedom,  we consider the broken phase in which the potential has a symmetry breaking  minimum at $|\phi| = v$. The plane wave solutions 
 to the linearized Maxwell equation (\ref{gauss})  with 
$A_\mu \propto e^{ i k \cdot x}$ and 
$k^\mu = (\omega , \vec k)$  lead to the 
dispersion relation $\omega = \sqrt { |\vec k|^2 + m_{\pm}^2 }$ ,  where the  photon  masses are given by 

\begin{equation}\label{mass}
m_{\pm} = { \pm (\kappa + e v^2 g) +
\sqrt { (\kappa + e v^2 g)^2 + 4 e^2 v^2 ( 1 - v^2 g^2/4) }
\over 2  ( 1 - v^2 g^2 /4)}  \, .
\end{equation}
The two values for the photon mass are related with two different 
polarizations of the electromagnetic wave.
 From the  plane wave solution for  $A^\mu $  and assuming that the wave propagates along the $x-$axis  we   find that the  the electric field  can be written as

\begin{equation}\label{pol}
\vec E \,\propto 
 \bigg( \pm {i m_{\pm} \over \omega } \, , \, 1 \bigg )
 e^{ i  k \cdot   x} \,  . 
\end{equation}
Hence the waves are  neither transverse nor longitudinal;  instead the solutions correspond to   right-handed  and left-handed longielliptic waves.
We notice that the two masses in (\ref{mass}) may also be deduced from the gauge propagator, or by the explicit analysis of the corresponding Maxwell-Proca equation. 

From the  work of Pisarski and Rao \cite{pisarski} it is known 
 that the combined effect of the  CS term with the mass induced by the Higgs mechanism produces two gauge modes with different masses.  
From  (\ref{mass})  it follows that   to have two distinct masses it is required that  both spontaneous symmetry breaking  
and at least one of the ${\cal P}$ and ${\cal T}$ violating terms exist.
 The induced term $e v^2 g$  simply adds to the CS mass $\kappa$,
as we are considering  here the topologically trivial  sector of the theory.

 Our result in (\ref{mass}) reduces to well known cases when the corresponding 
limits are considered: 
$(i)$ If the parity violating terms vanish $(\kappa = g = 0)$ we obtain the usual gauge  mass $m_{\pm} = ev$ produced by the spontaneous symmetry breaking; 
$(ii)$ In the symmetric phase $v = 0$ there is only one propagating mode with
mass $\kappa$. The existence of  massive gauge invariant theories in $(2 + 1)$ dimension,  without the Higgs mechanism, is known from the topological massive gauge theories \cite{deser}; 
$(iii)$ If we cancel the magnetic moment $(g = 0)$ we recover the result of
Paul and  Khare \cite{paul1}. 

The  stability  of the model is a genuine concern, we now address this point. In terms of the electric and magnetic fields the energy density  $(T_{00})$ can be written as 
\begin{eqnarray}\label{t00}
   T_{00}     =     {1 \over 2}  \left( 1 - {g^2 \over  4 }  |\phi|^2 \right)
 \left(   E _i^2   +  B^2 \right) 
    +      {1 \over 2}   | \nabla_0  \phi  |^2  +  
    {1 \over 2}  | \nabla_i  \phi  |^2    +  V (\phi)   
   \, .   \end{eqnarray}
Hence  one may think that  the magnetic moment contribution  will in general  spoil
the positive definiteness of the Hamiltonian. 
 
To investigate  the conditions required to have  an stable model  consider 
small scalar and gauge field fluctuations around the vacuum solution, in such a way that only quadratic terms on the fluctuation fields  are  retained.  
First, it is  straightforward to check that the model is stable  for small fluctuations of the  scalar field. This fact would not be true in the presence of a constant  external  magnetic field; however we recall that, because of the   topological mass term,  a constant  magnetic field is not  a solution of the equations of motion.
The next step is to  consider gauge field fluctuations,  whereas the scalar field remains at the minimum of the potential, $i.e.$ $|\phi| = v$.  Clearly the energy  remains positive definite if the following relation  holds

\begin{equation}\label{condi1}
{4 \over g^2} \geq v^2 \, .
\end{equation}
If  the previous relation is not satisfied then the model is no longer positive definite.  In this case any gauge field fluctuation with wave vector  $\vec k$, such that 
\begin{equation}\label{c0ndi2}
|\vec k | > {v^2 e^2 \over g^2 v^2 - 4} - {m_{\pm}^2 \over 2}
\end{equation}
will render the model unstable.  

For the purposes of our pursuit  we  have considered  that  condition (\ref{condi1}) holds, 
thus expect the theory to  be  stable both for  classical and quantum 
fluctuations of the  fields. In particular,   
for a $\phi^2$ potential  where  $v = 0$  condition
(\ref{condi1})  is   satisfied.   The  analysis above  considered the stability 
of the model under the  assumption of small field fluctuations. The analysis of the stability under large field fluctuations  is a matter of further work.
In this section we have  analysed  the propagating modes  and the stablity of these  solutions in the topological trivial sector ot the theory; the stability of the soliton vortex configuration will be discussed at the end of section \ref{rota}.

\section{The pure Chern-Simons limit}\label{pure}\indent

A theory in which the Maxwell kinetic term is dropped and the dynamics is solely 
provided by the CS term  has been frequently considered \cite{deser,review2}.  As   mentioned in the  introduction,  charged vortex solutions   are possible in these models. 
In  the Maxwell-Chern-Simons theory without magnetic moment interactions the pure CS limit can be  formally 
obtained by taking the limit $e^2$, $\kappa$ $\to \, \infty$   keeping
$e^2 /\kappa$ fixed and rescaling the gauge field by $A_\mu \, \to A_\mu/e$ \cite{boya}. This limit is shown to be equivalent to simply neglecting the $F_{\mu\nu}^2$ term in the lagrangian.
In  the model described by (\ref{lag})  
there is a particular relation between the CS mass and the anomalous
magnetic moment for which the  equations(\ref{eqm2}) for the gauge fields reduce 
from second- to first-order differential equations  \cite{stern,torres,latinsky}, 
similar to those of the   pure CS type.
To obtain this  limit  we   notice that  if we set 

\begin{equation}\label{kappag}
 \kappa = - {2 e \over g}
\end{equation}
then   any solution to the equation (\ref{eqm2}) can be written as

\begin{equation}\label{eqcs1}
 F_\mu  = {1 \over \kappa} J_\mu  \, + \, \lambda G_\mu
  \,    ,   
\end{equation}
where $\lambda$ is an arbitrary constant and  $G_\mu$ is solution of the homogeneous field equation 

\begin{equation}\label{eqm3}
 \epsilon_{\mu \nu \alpha}  \partial^\mu  G^\alpha 
    =   - \kappa G_\nu    \,  .
\end{equation}

In the vacuum the original theory possesses   two gauge propagating modes with  different masses. In the limit  $ \kappa = - {2 e / g}$ the  gauge field masses in (\ref{mass})  reduce to 
$m_+ = \kappa e^2 v^2 / (\kappa^2 - e^2 v^2) $  and $m_- =\kappa$ respectively.  Clearly the excitations described by the field $G_\mu$  carry a mass $m_- =\kappa$.
However if  the nonperturbative sector with vortex solutions is considered,  
it is easy to  see that any solution  to the homogeneous equation (\ref{eqm3})
gives an infinite contribution to the energy of the configuration.
Consequently   for the solitons to have finite energy  
 the constant $\lambda$   in (\ref{eqcs1})  should  be set   to zero.
Hence  when the   relation (\ref{kappag}) between the coupling constant holds,  the 
gauge field equation of motion  reduces to  

\begin{equation}\label{eqcs2}
 F_\mu  = {1 \over \kappa} J_\mu  \, .  
\end{equation}
This is a first order differential equation that  automatically satisfies the 
second order equation (\ref{eqm2}) when  $ \kappa = - {2 e / g}$.
This  equation has   the same structure as   that of the pure CS
 theory \cite{deser,review2}. So  we  shall refer to Eqs.   (\ref{kappag}) and (\ref{eqcs2}) as the pure CS  limit. However  we should notice that   
the explicit expression  for $J_\mu$ differs from that of the usual  pure CS  theory, because according  to  Eq. (\ref{dercov}) and  Eq. (\ref{cur1})  $J_\mu$ receives contributions from the anomalous magnetic moment.

The gauge field equation  (\ref{eqcs2})  represents a first order differential equation, so its propagating modes should be characterized by only one mass.
It is straightforward to check that  this mass is given by  $m_+ = \kappa e^2 v^2 /(\kappa^2 - e^2 v^2)$.  The other  mass $m_- = \kappa$ decouples  in the non-trivial topological sector due to the  finite energy condition. 
Henceforth  we  work on  the  limit in which Eqs. (\ref{kappag}) and (\ref{eqcs2})  are  valid.  Consequently we  shall consider  Eq. (\ref{eqcs2}) instead of the Eq. (\ref{eqm2}) as the equation of motion for the gauge fields.   

In terms of the    gauge  invariant potential  in (\ref{ainv}) the pure CS equation  (\ref{eqcs2}) reads

\begin{equation}\label{eqcs3}
F_\mu = -  { \kappa e^2 |\phi|^2  \over    \left( \kappa^2 - e^2 |\phi|^2 \right)} 
 \tilde{A}_\mu   \, .  \end{equation}

\section{ Non-topological solitons}\label{nontopo}
\subsection{Bogomol'nyi limit }\label{bogo}\indent

 In the    Bogolmol'nyi limit    all  equations of motion are known to become 
first-order differential equations \cite{bogo};  furthermore  it is possible  to recast  
the equations of motion as self-duality  equations. 
In the pure CS limit  (\ref{kappag}) the gauge field equations  have been  already reduced to the  first-order equations   in (\ref{eqcs2}); however the scalar field  is still governed by the second-order equations  (\ref{eqm1}). 
Here we outline the necessary steps to  derive  the self-duality limit. The self-duality 
equations for the $\phi^2$ model have   been reported previously 
\cite{tlee,torres2} but are mentioned   here for completeness and also to be used in the discussion of  domain wall  solutions. 
We can exploit the pure CS equation (\ref{eqcs3}) to eliminate $A_0$ 
and $E_i$ from the  expression of the energy density in  (\ref{t00}) . Hence,  we can write down   the energy $E = \int d^2x T_{00}$  for a static configuration as

\begin{equation}\label{ener1}
E  = \int d^2x \left( {  \kappa^2 - e^2 |\phi|^2  \over 2 e^2  |\phi|^2}   B^2 
+ {1 \over 2} \left[  | \partial_i \phi |^2  +
 { \kappa^2  e^2  |\phi|^2    \over  \kappa^2 - e^2|\phi|^2   } 
  \tilde{A}_i^2
  \right]  + V(\phi)   \right)     \, . 
\end{equation}
To ensure the positivity of the energy
here it is assumed that  $\kappa^2 \geq e^2 |\phi|^2$ .
The energy written in the previous  form is similar to the expression that appears in  the Nielsen-Olesen model. Thus,   starting from Eq. (\ref{ener1}) we  can  follow  the usual  Bogomol'nyi-type arguments in order to  obtain the self-dual limit.  The energy may then  be rewritten, after an integration by parts, as 

\begin{eqnarray}\label{ener2}
E    &= &   {1\over 2}   \int d^2x \left[  
 {  \kappa^2 - e^2 |\phi|^2 \over  e^2  |\phi|^2}
 \left( B  \mp  { \kappa  e  |\phi|^2 \over \sqrt{\kappa^2 - e^2 |\phi|^2} }  \right)^2
+  \left( \partial_{\pm}  |\phi|   - i  { \kappa  e  |\phi|  \over \sqrt{\kappa^2 - e^2 |\phi|^2} } 
 \tilde{A}_{\pm} \right)^2 \right] 
            \cr  
            &+ & \int d^2 x  \left[ V(\phi) - {1 \over 2} \kappa^2 |\phi|^2   \right]
\,  +  \, {\kappa^2 \over e} |\Phi|   \, , 
\end{eqnarray}
where   $\partial_{\pm} = \partial_1 \pm i \partial_2$,
 $\tilde{A}_{\pm} =  \tilde{A}_1 \pm i  \tilde{A}_2$ and 
$\Phi$ denotes the magnetic flux.
From the previous equation we observe that 
 the energy is bounded below; for a fixed value of the magnetic flux  the lower  bound is given by   $E \, \geq{\kappa^2 \over e}   \Phi$,  provided  that the potential  is chosen   as  $V(\phi) =  {m^2 \over 2} |\phi|^2$ with the critical value 
$m = \kappa$, $i.e.$ when the scalar and 
the  topological masses are equal. Therefore   in this limit  we are necessarily  within the  symmetric phase of the theory. From Eq. (\ref{ener2}) we  see that the lower bound for the energy 

\begin{equation}\label{ener3}
 E \, = \, {\kappa^2 \over e} | \Phi| \, = \, {\kappa \over e} |Q|
  \,  , 
\end{equation}
is saturated when the following self-duality equations are satisfied

\begin{equation}\label{sd1}
   B \,   =  \, \pm {\kappa e |\phi|^2 \over \left[ \kappa^2  - e^2  |\phi|^2  \right]^{1/2}}
  \,  , \end{equation}

\begin{equation}\label{sd2}
 \!\!\!\!\!\!\!\!  {1 \over 2} \partial_{\pm} |\phi|^2  \,  =   \,  
{i e   \kappa |\phi|^2 \over \left[ \kappa^2 - e^2 |\phi|^2 
\right]^{1/2}  }\,   \tilde{A}_{\pm}          \,  ,
\end{equation}
where the  upper (lower) sign corresponds to a positive (negative) value of the magnetic  flux.  
We should remark that the   present model in the self-duality limit   corresponds  to the bosonic part of a theory with an $N+2$ extended supersymmetry \cite{navra}. 

Equation (\ref{sd1})   implies  that the magnetic field vanishes whenever $\phi$  does.   The finiteness energy condition forces   the scalar field to vanish  both at the center of  the vortex (except for the nontopological solitons, see next section)  and also at spatial infinity;   consequently   the  magnetic flux of the vortices  lies in a ring.  It is interesting to observe  that  (\ref{sd2}) can be written as an explicit  self-duality equation;
indeed if we define a new covariant derivative as 
$\tilde{D}_i = \partial_i - i  2 e \kappa/\sqrt{\kappa^2 - e^2 |\phi|^2}$,  then Eq. (\ref{sd2}) 
is equivalent to 

\begin{equation}\label{sd3}
   \tilde{D}_i   |\phi|^2  \,  =   \,  \mp i   \epsilon_{ij} \tilde{D}_j  |\phi|^2          
  \,  .
\end{equation}

Equations   (\ref{sd1}) and   (\ref{sd2}) can be reduced
to one nonlinear second order differential  equation  for a unknown function. 
To do this  first notice that   (\ref{sd2})
implies  that  $ \tilde{A}_i$ can be determined in terms  of the scalar field,
on substituting the result in  (\ref{sd1})  we get 
  
\begin{equation}\label{neq1}
  \partial_i \left[ { \sqrt{\kappa^2 - e^2 |\phi|^2}  \over 2 e \kappa} \partial_i 
 \ln{\left(|\phi|^2  \right) }  \right]    + 
{\kappa e |\phi|^2 \over \sqrt{\kappa^2 - e^2 |\phi|^2}}  \, = \, 0
\, .
\end{equation}

\subsection{Rotationally symmetric solutions}\label{rota}\indent

The self-dual limit  is attained  for a $\phi^2$ potential,  consequently topological solitons  do not exist. However   the theory allows nontopological soliton solutions.
In order to look for vortex solutions we nowconsider   static rotationally 
symmetric solutions of vorticity $n$ represented by the ansatz

\begin{eqnarray}\label{ansatz}
  \vec A (\vec \rho) &= &  - \hat{\theta} { a(\rho) - n \over e\rho}  \, ,
\qquad A_0(\vec \rho) = {\kappa \over e} h(\rho) \, ,  \nonumber\\
 \nonumber\\
   \phi(\vec \rho)  &=& {\kappa \over e} f(\rho) \exp{(- i n\theta)}    \,,    \end{eqnarray}
where $\rho, \theta $ are the polar coordinates.  After substituting this ansatz the self-duality Eqs. (\ref{sd1})  and (\ref{sd2}) become

\begin{equation}\label{eq1}
 {1 \over  \rho }{d a \over d \rho} = 
\mp { \kappa^2   f^2  \over (1 - f^2)^{1 \over 2} }
\, , 
\end{equation}

\begin{equation}\label{eq2}
{ d f  \over d \rho}= \pm { f a  \over \rho (1 - f^2)^{1 \over 2} }  \, , 
\end{equation}
 Notice that   the function  $h(\rho)$   can be explicitly solved using Eq. (\ref{eqcs2}) as  $h(\rho) = \left( 1 - f^2 \right)^{1/2}$. 
In what follows,  we  select the signs (upper signs in  the previous equations)
corresponding to positive magnetic flux $(n  >  0)$. The equations for $n < 0$
are obtained  with the replacement $ a \to - a$, $f \to f$ and  $h \to - h$.

The   boundary conditions   are selected  in such a way that the fields are 
non-singular at the  origin and give  rise to a finite energy solution.
The first condition implies that  $a(0) = n$ and $n \, f(0) = 0$. 
Whereas  the finiteness of the energy  implies that 
 $a \to  -\alpha_n$ and $ f\to 0$ as  $\rho  \to \infty$. 
Notice  that these requirements leave   $a(\infty) = - \alpha_n$ undetermined. Consequently  the magnetic flux for  nontopological solitons  is not quantized, 
but rather   is a continuous parameter describing the solution.    Indeed
once that the boundary conditions are known,  the ``quantum'' numbers of the soliton can be  explicitly computed.  With the ansatz (\ref{ansatz}) the magnetic field is $B= (1/r) (d a / dr)$, and  so using the boundary conditions   the magnetic flux  and electric charge are  

\begin{equation}\label{bflux}
\Phi \, = \, - {Q \over \kappa} \, = \,   \int d^2x B = {2 \pi \over e }  \left( a(0) - a(\infty)  \right) =
 {2 \pi \over e }  \left( n + \alpha_n  \right)
  \,   . 
\end{equation}
We shall later see that  for every value of the vorticity $n$ the allowed values for $\alpha_n$ are restricted according to Eq. (\ref{bounds}). The solutions are also characterized by the spin $S$ (which  in general  is fractional) and the   magnetic moment $M$. 
The spin   is obtained  from   the 
gauge invariant symmetric energy-momentum tensor (\ref{tmunu})  via  
 $S = \int d^2x \left( \epsilon^{ij} x_i T_{0j} \right) $; whereas  for the 
magnetic moment we use  Eq. (\ref{magne2}).
An explicit  calculation  yields 

\begin{eqnarray}\label{topnumbers}
 S &=& {\pi \kappa \over e^2} \left( \alpha_n^2 - n^2  \right)      \, , \nonumber\\
  M &= &   - {\pi \over e}  \int_{0}^{\infty}{r^2 {dh \over dr} dr} 
   \,.    \end{eqnarray}
Notice that the magnetic flux, the charge and the spin can be  explicitly integrated,
because they  depend solely  on the boundary conditions. Instead,  the magnetic moment depends on the structure factor of the vortex configuration.

For the rotationally  symmetric  ansatz (\ref{ansatz}) the equation
(\ref{neq1})  reduces to 

\begin{equation}\label{neq2}
{1 \over \rho} {d \over d \rho}  \left[ \rho { d f \over d \rho }
\right]  = { 1 \over f ( 1 - f^2)}
 \left[ \left( {d f \over d \rho } \right)^2 - \kappa^2 f^4 \right] \, .
\end{equation}
The same result is obviously obtained if we combine Eqs. (\ref{eq1}) and (\ref{eq2}). If we consider the case of small $f$ we can approximate
$ ( 1 - f^2)^{ -1} \approx 1$. Then,    Eq. (\ref{neq2}) 
reduces to the  rotationally symmetric form of  the Liouville's  equation,
which has the following solution

\begin{equation}\label{lio}
 f(\rho) = {2 N \over  \kappa \rho } \left[ 
\left( { \rho \over \rho_0} \right)^N + \left( { \rho_0 \over \rho} \right)^N
\right]^{-1}
\end{equation}
where $N$ and $ \rho_0$ are arbitrary constants.

As mentioned before, the  finiteness of the energy implies that 
$ f (\infty) = 0$, and therefore the value $ a ( \infty) = - \alpha_n$
is not constrained. We  asymptotically  solve equations 
(\ref{eq1}) and  (\ref{eq2}) as $\rho \to \infty$, 

\begin{eqnarray}\label{large}
 f (r) &=& {C_n \over  (\kappa \rho)^\alpha} -
{ C_n^3 \over 4 (\alpha - 1)^2 \,  (\kappa \rho)^{3 \alpha - 2}} +
O  \left(  (\kappa \rho)^{- 5 \alpha + 4} \right)  \, , \nonumber \\
a (r) &=& - \alpha + 
{C_n^2 \over 2 (\alpha - 1) \,  (\kappa \rho)^{2 \alpha -2}}
- O \left( (\kappa \rho)^{- 4 \alpha + 4} \right)
\, ,  \end{eqnarray}
where $\alpha \equiv\alpha_n$ and $C_n$ is a constant. Notice that  $f(\rho)$ is asymptotically small so the first 
two terms of the previous expansion for $f(\rho)$  can be directly obtained
from the Liouville approximation (\ref{lio}) if we set 
$N = \alpha - 1$ and $( \kappa \rho_0)^{\alpha - 1} = C_n/ 2 (\alpha - 1) $.
 
In the origin the boundary conditions are   $a(0) = n$ and $n \, f(0) = 0$.
Hence, it is convenient to consider separately   two categories of solutions: 
$(i)$ the zero vorticity  and $(ii)$ the 
nonzero vorticity. 

\vskip0.8cm
\centerline{{\bf  ({\it i}) $n = 0$. Nontopological solitons:}}

In this case   $a(0)$ must
vanish to insure that  the solution is non-singular at the origin, but 
$f(0) = f_0$ is not so constrained. These are nontopological 
solitons that  are characterized
by the value of the magnetic flux $ \Phi = {2 \pi \over e} |\alpha_0|$.
The large-distance behavior is given 
by Eqs. (\ref{large}), while as $\rho \rightarrow 0$ we obtain a 
power-series solution

\begin{eqnarray}\label{sh1}
 f (\rho ) &=& f_0 - {f_0^3 \over 4 (1 - f_0^2)}  (\kappa \rho)^2 +
{f_0^5 ( 4 - f_0^2) \over 64 (1 - f_0^2)^3}  (\kappa \rho)^4
+ O \left(  (\kappa \rho)^6 \right) \, , \nonumber \\
a (\rho)&=& - {f_0^2 \over 2 ( 1 - f_0^2)^{1 \over 2}} (\kappa \rho)^2
+ {f_0^4 (2 - f_0^2) \over 16 (1 - f_0^2)^{5 \over 2} } (\kappa \rho)^4
+ O \left( (\kappa \rho)^6 \right) \, . 
\end{eqnarray}
Acceptable soliton solutions exist for values of $f_0$ in the range
$ 0 < f_0 < 1$. The short- and  large-distance behaviors  of the solutions
are related, since $\alpha_0$ is a function  of $f_0$, see Fig. 1.  
The extreme values on this plot  are obtained as follows.
If    $f_0 \ll 1$ then  $f(\rho)$  remains  small for all $\rho$  and can therefore be approximated by the 
Liouville solution. Comparing the   expansion  in  (\ref{sh1}) 
with the one obtained from  the Liouville solution  near the origin we see that the 
constant $N$ in   (\ref{lio}) should be set to $1$ while 
 $\kappa \rho_0 = 2/ f_0 $. But now  the same  Liouville solution is also applicable in the large distance region, comparing with Eq. (\ref{large}) we obtain $\alpha = 2$.
Instead   as $f_0 \rightarrow 1$ we
find by numerical integration that  $\alpha_0 \rightarrow 1.755$. 
Thus,  
the magnetic flux varies continuously between $ \Phi = 0.877 ( 4 \pi /e)$
and  $ \Phi =  4 \pi /e$.  In  Fig. 2 we show profiles of the magnetic field $B$ as function of $\kappa \rho$ for several values of the parameter $f_0$.  For nontopological solitons the magnetic field decreases monotonically from its maximum value at the origin, so the soliton has a flux tube structure. 

\vskip1.5in
\let\picnaturalsize=N
\def\picsize{3.0in}
\def\picfilename{fig1.eps}
\ifx\nopictures Y\else{\ifx\epsfloaded Y\else\input epsf \fi
\let\epsfloaded=Y
\centerline{\ifx\picnaturalsize N\epsfxsize \picsize\fi \epsfbox{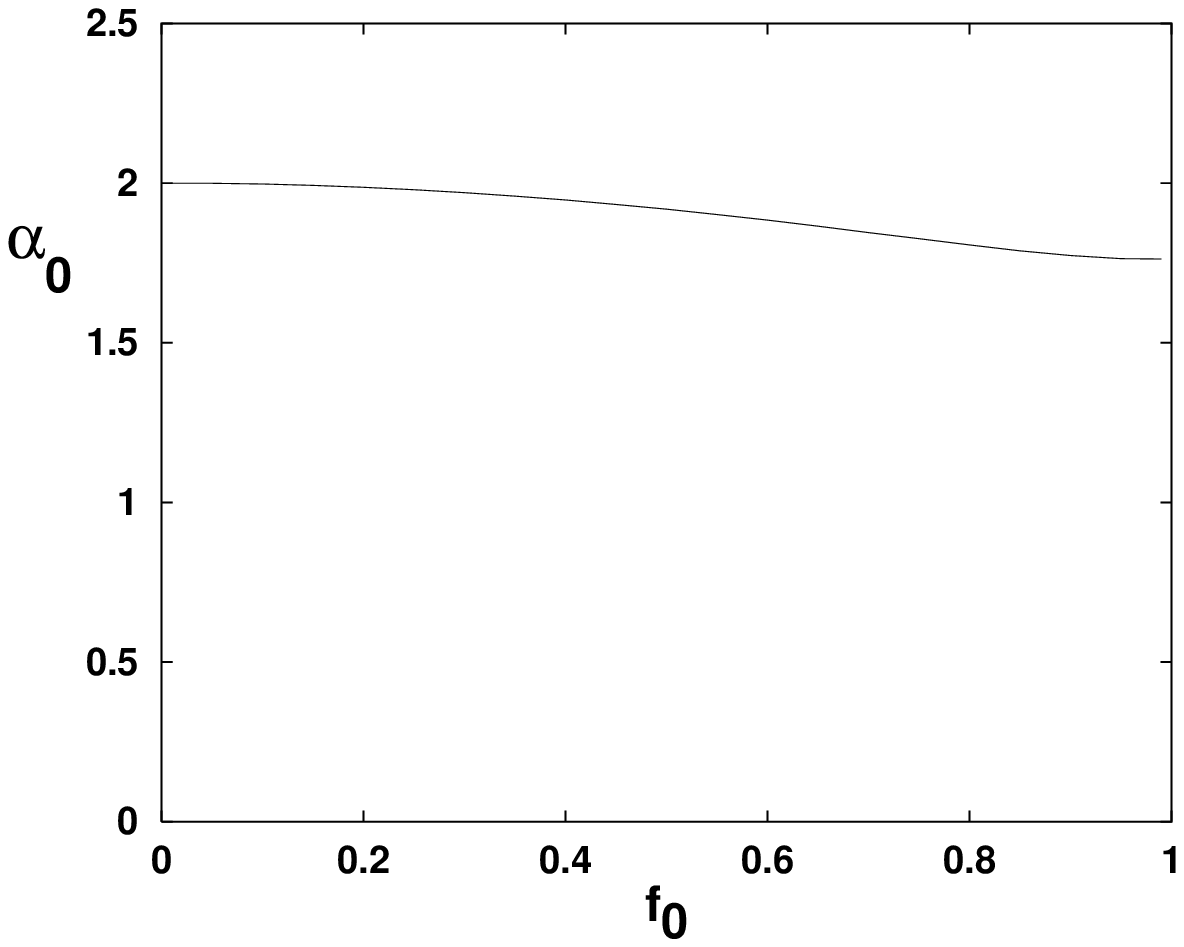}}}\fi
{\bf Fig. 1 
Behavior of   $\alpha_0$ as function of $f_0$ in  the case of 
 nontopological    solitons $(n=0)$.}

\bigskip

\let\picnaturalsize=N
\def\picsize{3.0in}
\def\picfilename{fig2.eps}
\ifx\nopictures Y\else{\ifx\epsfloaded Y\else\input epsf \fi
\let\epsfloaded=Y
\centerline{\ifx\picnaturalsize N\epsfxsize \picsize\fi \epsfbox{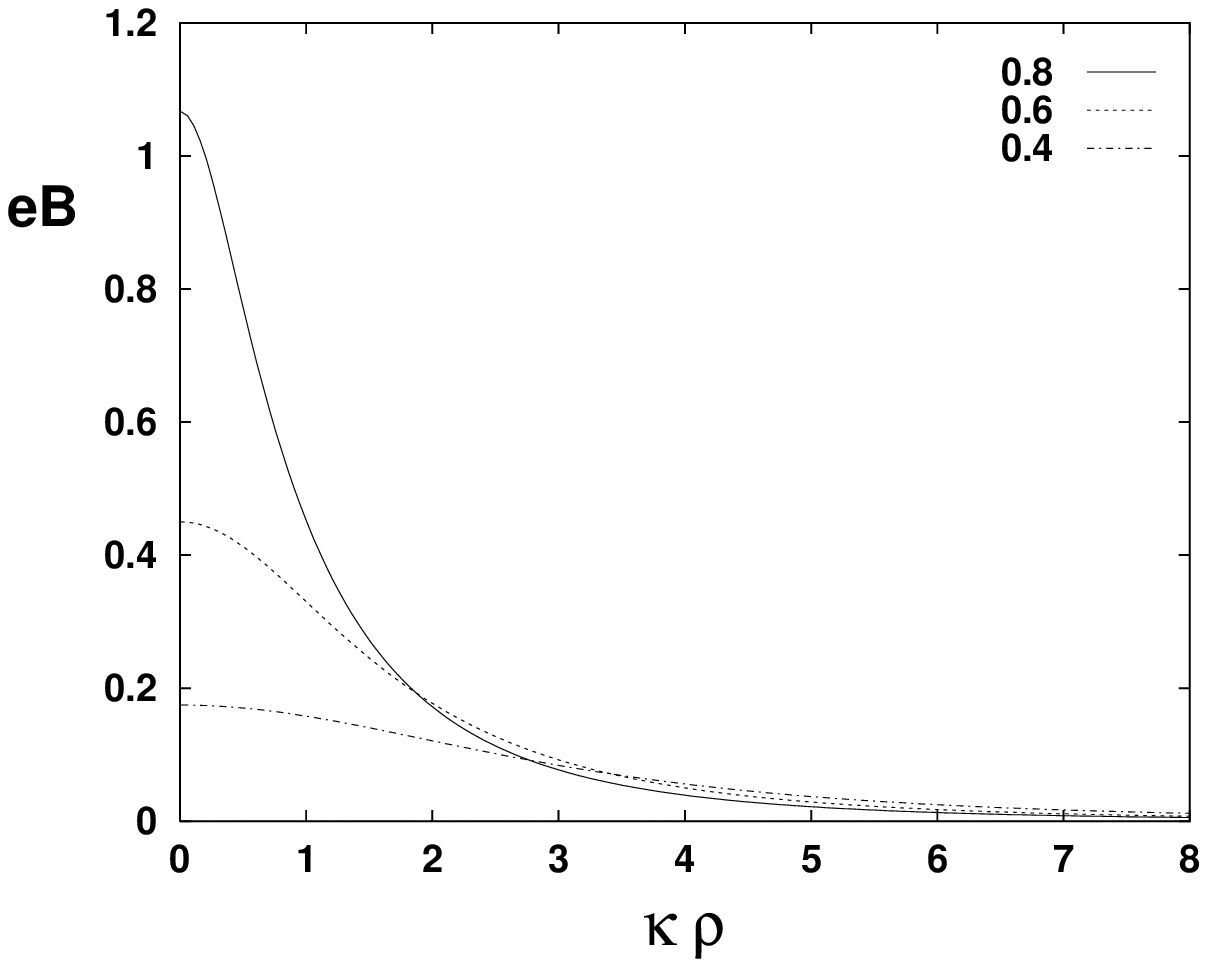}}}\fi
{\bf Fig. 2 
The magnetic field in units of $1/e$   for the  nontopological soliton solution 
 and values  of  the parameter $f_0$ of $0.8$, $0.6$ and $0.4$.}

\bigskip

\vskip0.8cm
\centerline{{\bf  ({\it ii}) $n  \neq 0$. Nontopological vortices:}}

Following  Jackiw $et. al.$ \cite{jack2}  we shall refer to  these configurations with non-vanishing vorticity  as  nontopological vortices. 
In this case the boundary conditions imply  that  $f(0)$ must vanish 
and $a (0) = n$. The large distance  behavior of the fields is given  by Eqs. (\ref{large}).   
For small $\rho$ a power-series solution  gives

\begin{eqnarray}\label{sh2}
f (\rho)  &=& f_n (\kappa \rho)^n   - {f_n^3 \over 4 (n +1)^2 } (\kappa \rho)^{3 n + 2} + O \left( (\kappa \rho)^{5 n + 2 } \right)   \, ,  \nonumber \\
 a (\rho) &=& n - {f_n^2 \over 2 (n +1)} (\kappa \rho) ^{2 n + 2} 
 + O \left( (\kappa \rho)^{2  n + 4} \right) \, .
 \end{eqnarray}

The constant $f_n$ is not  determined by the behavior of the field 
near the origin, but  is instead fixed  by requiring proper behavior at spatial infinity
and   that the function remains  real for all $\rho$.
For each integer $n$ there will be a continuous set of solutions
corresponding to the range  $ 0 < f_n < f_n^{max} $.
For values such that $f_n > f_n^{max}$ there are no real  solutions to the field 
equations  (\ref{eq1})  and  (\ref{eq2}),
because the condition $f (\rho)< 1$ is not satisfied for all $\rho$. 

If we consider the case $ f_n \ll 1$,  then   $f(\rho)$ is small for all $\rho$ and the 
Liouville  approximation can be used everywhere.  In order to match the solution
in  (\ref{lio}) with both the   short- and large-distance approximations in  (\ref{sh2}) and (\ref{large})   we should set:
 $N = n + 1$ and $(\kappa \rho_0)^{(n + 1)} = 2 (n + 1) /f_n$ and  
 $\alpha_n = n + 2$.
Hence the corresponding value of the magnetic flux  is   $ \Phi =  4 \pi (n + 1)/e$

The value   $\alpha_n = n + 2$ is  an upper bound.
On the other hand, as   $ f_n \rightarrow f_n^{max}$ we find 
that $\alpha_n$ tends to a minimum value $\alpha_n^{min}$. 
In fact, it is possible to derive sum rules  \cite{escalona} to prove 
that   $\alpha_n$  is restricted as, 
$ n \,  < \, \alpha_n  \, < \, n + 2 $. However, from a numerical analysis 
we find a more stringent condition on the lower  bound. 
In table ~\ref{tabla} we present the values of the parameters $\alpha_n^{min}$ and    $ f_n^{max}$  for several  vorticity numbers $n$.  We observe that the lower bound  for $\alpha_n$ can be taken as $n +1 $, this approximation improves  for larger $n$. 
Hence, we conclude that  the parameter $\alpha_n$ satisfies the 
inequalities 

\begin{equation}\label{bounds}
 n + 1  \,  < \, \alpha_n  \, < \, n + 2 \, .
\end{equation}
According to the previous results we can set $\alpha_n^{min}  \approx n + 1$.
Thus,     for each integer $n$ the flux varies continuously between 
$\Phi_n^{min} = {4 \pi \over e} \left[ n + {1 \over 2} \right]$  
and  $\Phi_n^{max} = {4 \pi \over e} \left[ n + 1 \right] $.  Similarly
(see  Eq. (\ref{ener3}))  the energy spectrum   consists of bands of finite width:

\begin{equation}\label{edesi}
 {4 \pi  \kappa^2 \over e^2} \left[ n + {1 \over 2} \right]  \leq E_n \leq
 {4 \pi  \kappa^2 \over e^2} \left[ n +  1 \right] \, . 
\end{equation}

In  Fig. 3 we show the magnetic field $B(\rho)$ for the $n=1$ and $n=2$ solutions. For nontopological vortices the magnetic field is localized within a ring.  

If we select the value of the parameter $f_n = f_n^{max}$  in (\ref{sh2}), 
 the function $f(\rho)$ will reach it maximum value at a given 
radius $\rho = R_n$,  $i.e.$  $f(R_n) = 1$. This  parameter $R_n$ can be considered 
as the radius of the soliton.  In Fig. 4 we show the profiles of the energy density as a function of $\kappa \rho$
for several  values of $n$. The energy is indeed concentrated in the region 
$\rho \approx R_n$. 
 In the next section we shall see   that in the large $n$ limit,  
 the vortex  can be   considered as a 
 ring  of radius  $ R_n \approx n / \kappa$ and thickness $1 / \kappa$.
Furthermore, in the region $\rho \sim R$  the fields can  be correctly approximated  by 
 a domain wall solution.   The value of $f_n^{max}$ as a functions of $n$ is plotted in Fig. 5,  where the solid line indicates the prediction  of  Eq. (\ref{coex}) which we obtain in the next 
section using the domain wall approximation.

\let\picnaturalsize=N
\def\picsize{3.0in}
\def\picfilename{fig3.eps}
\ifx\nopictures Y\else{\ifx\epsfloaded Y\else\input epsf \fi
\let\epsfloaded=Y
\centerline{\ifx\picnaturalsize N\epsfxsize \picsize\fi \epsfbox{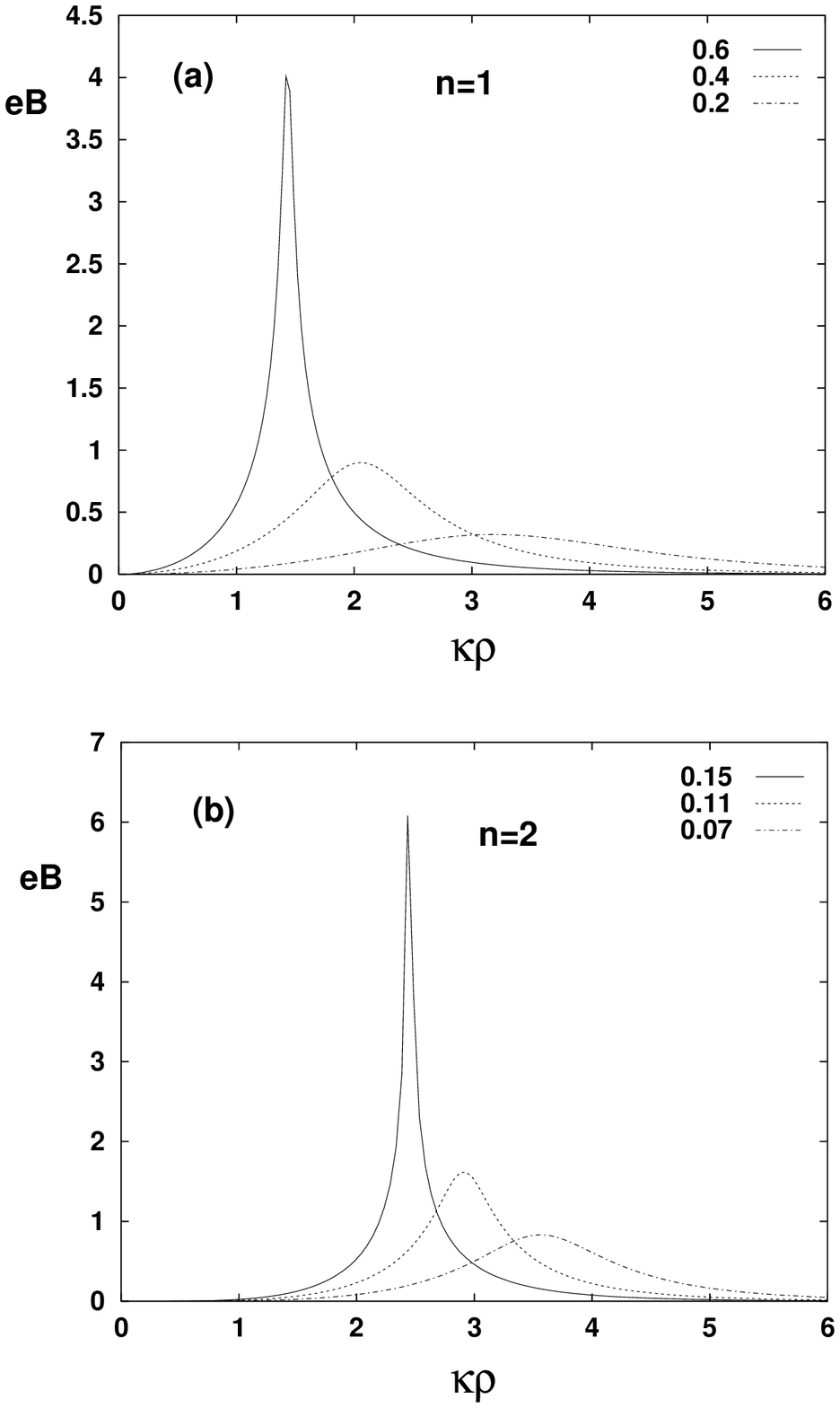}}}\fi
{\bf Fig. 3
The magnetic field in units of $1/e$.  (a)    For the  nontopological vortex  
solution  with $n=1$ and values  of  the parameter $f_1$ of  $0.6$,  $0.4$ and 
$0.2$.    (b) For the $n=2$ vorticity solution with  values of $f_2$ of 
$0.15$, $0.11$ and $0.07$. }

\vskip1.0cm
\let\picnaturalsize=N
\def\picsize{3.0in}
\def\picfilename{fig4.eps}
\ifx\nopictures Y\else{\ifx\epsfloaded Y\else\input epsf \fi
\let\epsfloaded=Y
\centerline{\ifx\picnaturalsize N\epsfxsize \picsize\fi \epsfbox{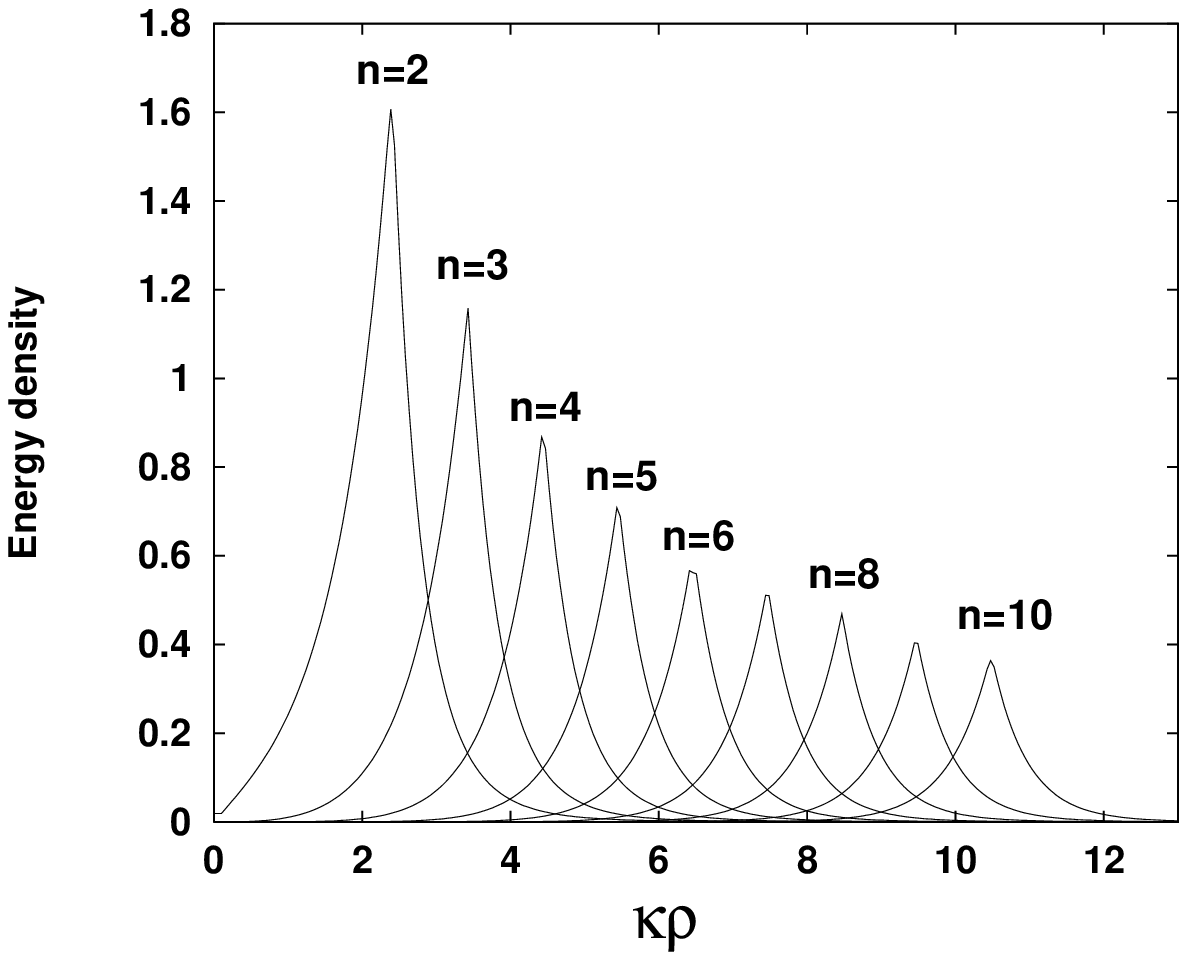}}}\fi
{\bf Fig. 4
Plots of   energy density for the vortex solutions with several   values of
the vorticity  number $n$.  In all of  these configurations the maximum value of 
the    parameter  $f_n=f_n^{max}$  was selected from table  I. }

\bigskip

\vskip1.0cm
\let\picnaturalsize=N
\def\picsize{3.0in}
\def\picfilename{fig5.eps}
\ifx\nopictures Y\else{\ifx\epsfloaded Y\else\input epsf \fi
\let\epsfloaded=Y
\centerline{\ifx\picnaturalsize N\epsfxsize \picsize\fi \epsfbox{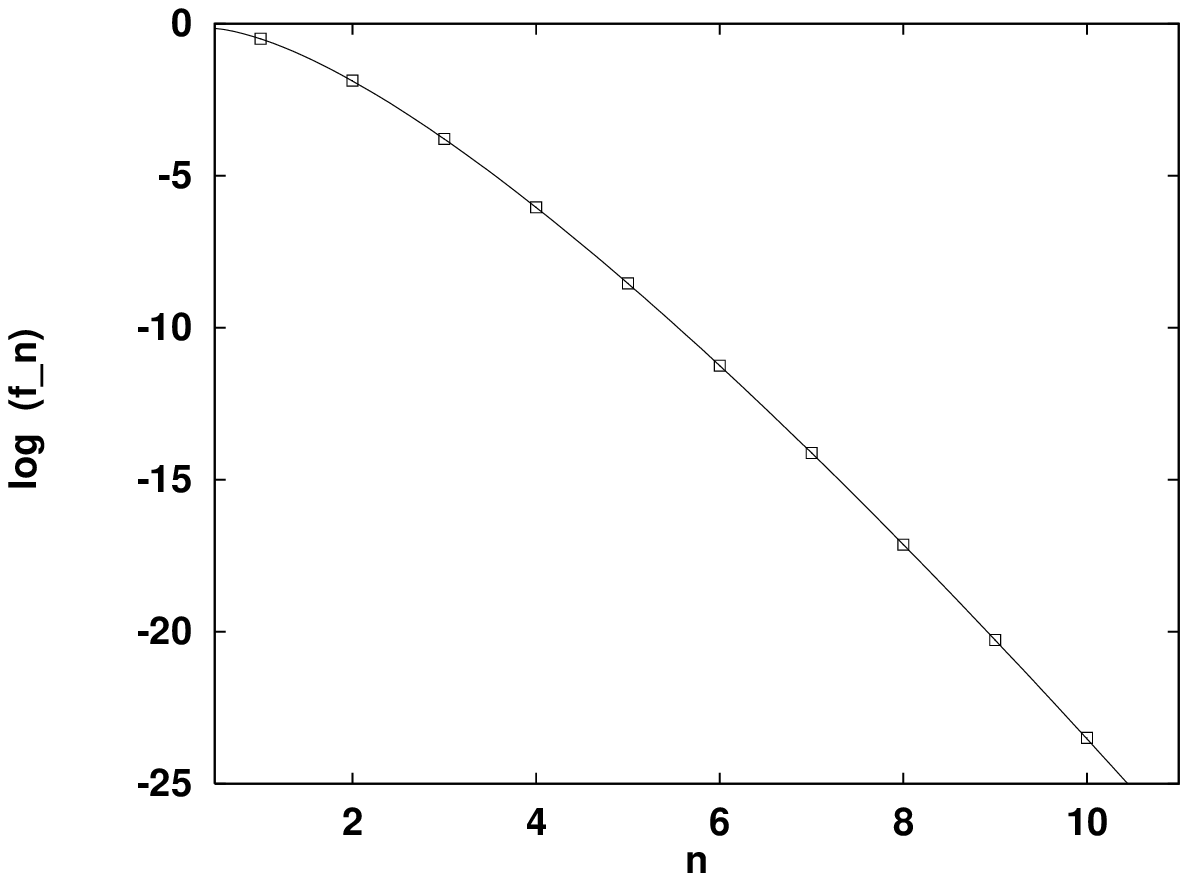}}}\fi
{\bf Fig. 5
Behavior of  $f_n^{max}$ as a function of $n$. The solid line corresponds to the asymptotic  formula (6.14), while the squares represent  actual data.}

\vskip1.0cm

We conclude this section with some comments about the stability 
of the   vortex solutions and also about the  interaction between vortices.
The vortices are neutrally stable at the self-dual  point $ m = \kappa$. This  fact is  easily demonstrated on account of  the  relation  (\ref{ener3}) between the  energy and the charge:  $E =\kappa |Q|/e$:   
The mass of the elementary excitations of the theory (scalar particles)
is $m$ and the charge $e$; due to charge conservation,  a 
decaying soliton should radiate $Q/e$ ``quantas'' of the scalar
particles, thus  the energy of the elementary excitations will be
 ${\cal E} = m Q/e$. This indicates that the  vortices are at
the threshold of stability against decay to the elementary 
excitations because   the ratio
$  {E_n / {\cal E}} = {\kappa / m}\, $
is  equal to one at the critical point $m = \kappa $. In fact it is possible 
to consider  a perturbative  method   away from the self-dual point  \cite{escalona}
to prove that the soliton is stable against dissociation into free scalar particles
when the scalar mass is bigger than the topological mass ($i.e.$ $m > \kappa$).
Instead  for $m < \kappa$ the soliton  becomes unstable.

The self-dual point also corresponds to a point in which 
 the vortices become noninteracting. 
Again this  property ensues   from  the fundamental relation  (\ref{ener3}).
Consider  two solitons of charges $Q_1$ and $Q_2$ of the same sign  that are far   apart. According to  (\ref{ener3}) their total energy is 
$E = {\kappa \over e} (Q_1 + Q_2)$. If   the two vortices
are superimposed at the same point, due to charge conservation  the
resulting configuration will represent a vortex solution of 
charge  $Q_1 + Q_2$. Then according to  (\ref{ener3})  the total energy
will  again be $E = {\kappa \over e} (Q_1 + Q_2)$. 
We therefore  conclude that the vortices are noninteracting. 
The perturbative analysis of  reference  \cite{escalona} shows that the
vortex-vortex interaction is repulsive if $m > \kappa$ and 
attractive if $m < \kappa $.
The self-dual point $m = \kappa$ represents a transition between 
a phase in which vortices attract and a phase in which they repel each other,
similar to the transition between type I and type II superconductors.  
In fact, what we  demonstrate  in the $\phi^2$ model 
 is that 
the attraction between vortices due to the interaction through the scalar 
field has the same strength as the repulsion due to the interaction 
through the vector field. Therefore when the range of the two interactions 
is the same ($m = \kappa$) the vortices become noninteracting. 
On the other hand if  the range of the scalar interaction is smaller
than the range of the vector   interaction
($ m > \kappa$) the inter-vortex potential is repulsive; while
for $ m  < \kappa$ the potential is attractive.

\section{Domain walls}\label{domwa}\indent

Domain walls  appear in   theories where the 
scalar potential  possesses two or more disconnected but degenerate minima.
 The field configuration  interpolates  between  two adjacent minima of the potential; the  infinitely long boundary  separating these  two vacua states  
is precisely  the domain wall. 
In $3 + 1$ dimensions the domain walls are planar structures, instead in $2 + 1$ dimensions  they correspond to  one dimensional structures with  finite energy 
 per unit  length.  Domain walls  solutions have been found in a  Chern-Simons  model \cite{jack2} with a scalar potential of the form
 $V(\phi) \propto |\phi|^2 (v^2 -  |\phi|^2)^2 $.

The  present $\phi^2$  theory possesses  a single minimum, yet  it is possible  to find  one dimensional  soliton solutions of the    domain wall type. Consider a one dimensional structure  depending only on   the $x$  variable,  both  at $x \to \infty$  and $x \to - \infty$ the scalar field  should vanish. However,  there can be  an intermediate region where  $\phi \neq 0$, $i.e.$, a region of false vacuum.
The  maximum of $\phi$ determines the position of the  wall.
In this  section  we show that such solutions indeed exist for the  
$\phi^2$ model.  The domain wall  carries both magnetic flux and electric charge 
per unit length. Furthermore,  these solutions  provide an approximate solution to the self-dual vortices
 in  the large flux limit (large $n$  limit). 

Seeking  a domain   wall solution parallel to the $y$-axis, the translational 
invariance of the theory implies that all the fields depend only on $x$.
By an appropriate gauge transformation the scalar field is made real everywhere
$ \phi = (\kappa/e) f$ and the potential $\vec A$ is selected along the $y$ axis.  Hence,  the  expression  (\ref{ener1}) for the energy  with a  potential 
$V(\phi) = {m^2 \over 2} \phi^2$  can be written as 
\begin{eqnarray}\label{ew1}
E = { 1 \over 2}  \int d^2 r \bigg[  
{\kappa^2 \over e^2} \left({d f \over d x} \pm m f   \right)^2
&+& \left({\left(1 - f^2  \right)^{1/2} \over f } {d A_y\over d x}
 \mp {\kappa f A_y \over \left(1 - f^2  \right)^{1/2}  }  \right)^2
\nonumber \\
&\mp& {m \kappa^2 \over e^2}{d f^2\over d x}
\pm \kappa {d  A_y^2\over d x}
\bigg]
\, . 
\end{eqnarray}
As mentioned earlier the boundary conditions for the scalar field
are $f(-\infty) = f(\infty) = 0$. The magnetic flux  per unit length $(\gamma)$
is given by  $\gamma = A_y(\infty)  - A_y(-\infty)$, so 
$A_y(\infty)  \neq A_y(-\infty)$  is required in order to get a non-vanishing magnetic flux.  A configuration is sought which   has a definite symmetry  with respect to the 
position $X$ of the domain wall, then    
$A_y(\infty)  = -  A_y(-\infty) \equiv \gamma/2$ is selected. 

The static solution is obtained  minimizing the energy per unit length  with 
$\gamma$ fixed. The boundary conditions cannot be satisfied if   the same upper (or  lower) signs  in  (\ref{ew1}) are used  for all    $x$.  Rather,  the upper signs in the region to the  right  of the domain wall ($x > X$) are selected, whereas  for
 $x < X$ we take the lower signs.
With this selection the minimum energy per unit length becomes 
\begin{equation}\label{ew2}
{\cal E} = {\kappa^2 m \over e^2 } f_0^2  +  { \kappa \over 4} \gamma^2 
\, , 
\end{equation}
where $f_0 \equiv f(X)$. This result is obtained provided that the fields
satisfy the following  equations:
\begin{eqnarray}\label{edw}
{df \over d x} &=& \mp m f \, \nonumber \\
{d A_y \over d x} = &=& \pm {\kappa f ^2 \over \left( 1 - f^2  \right) } A_y
\, , 
\end{eqnarray}
where the upper (lower) sign must be taken for $ x > X$  ($ x <  X$).
These equations  are easily integrated to give 
\begin{eqnarray}\label{sw}
f(x ) &=& e^{- m |x - X|}  \, ,  \nonumber \\
A_y(x) &=&  sgn(x - X) {\gamma \over 2}\left(1 - e^{-2 m |x - X|}  \right)^{\kappa/2 m }    \, . 
\end{eqnarray}

This is   a domain wall configuration   localized at   $x = X$  with a width 
of order $1/m$.  The solution to the  first equation in (\ref{edw}) does  not restrict the value   of $f_0$. However,  $f_0 = 1$ has to  set so 
the gauge field be continuous everywhere. The anti-kink configuration is obtained  by simply reversing   the signs of the fields in (\ref{sw}).

The domain wall carries a magnetic flux and charge per unit length
given by $\gamma $ and  $- \kappa \gamma$ respectively.
Although there  is a linear momentum  flow along the domain wall  given by 
\begin{equation}\label{lmw}
T_{0y} = {\kappa \over 2} {d A_y^2 \over d x} \, ,    
\end{equation}
where  Eqs. (\ref{tmunu})  and (\ref{eqcs3}) have been used, we notice that the flow at opposite sides of  the wall cancels, hence  the total linear  momentum  of the domain wall vanishes.  The  magnetic field is given by 
\begin{equation}\label{mfw}
B= {\kappa  \gamma \over 2}  e^{- m |x - X|} 
 \left(1 - e^{-2 m |x - X|}  \right)^{\kappa/2 m - 1} \, .
\end{equation}
Notice  that for $\kappa < 2 m$ the magnetic field is concentrated near 
$ x = X$ and falls off rapidly away from the wall. Instead for 
$\kappa  \geq 2 m$ the  magnetic field vanishes at  $x = X$ and
the profile  of $B$ is  doubled peaked  with maximums  
 at $x = X \pm {1 \over  m} | \ln \left({\kappa \over 2 m} \right)|$. 

The  method  presented in this section resembles the one 
used to derive the self-duality equations, so  it could be questioned wheter  
 the two methods are equivalent. 
In general it is not so: the  self-dual limit  is  valid when $m = \kappa$,
whereas for the  domain wall solution there is no such   restriction.
However, for   $m = \kappa$ it  is   straightforward to check that the fields in (\ref{sw})
exactly solve the self-duality equations (\ref{sd1})  and (\ref{sd2}), with the magnetic flux  and energy per unit length  determined  as:  
\begin{equation}\label{exflu}
\gamma = {2 \kappa \over  e }  \,,  \qquad {\cal E} = {2 \kappa^3 \over e^2}   \, .
\end{equation}

The domain wall solutions  in  (\ref{sw})  can also be adapted  to  approximate rotationally symmetric  configurations. 
Indeed,  the vortex configurations simplify in the large $n$ limit 
and  it is possible  to utilize the domain wall  as  an approximated  solution to  the self-dual  vortices. 
Let us consider a  r  ring  of  large radius $R$   and  thickness
of order $1 /m$  separating  two regions  of vacuum. 
The magnetic flux is concentrated
within this domain of  width $\sim  1/ m$   where   a region of false vacuum  ($\phi \neq 0$) is trapped. If   $R \gg1 /m \sim 1/\kappa$,  then the fields near the
ring should be well approximated by the domain wall solution (\ref{sw}). 
Nevertheless,  in  order to have a  configuration with vorticity $n$ the  phase of the scalar 
field should vary uniformly with angle, see  (\ref{ansatz});  hence  we   gauge transform the fields in  (\ref{sw}). Thus   in the region   $ \rho \sim R$
the fields  configuration   reads  
\begin{eqnarray}\label{vorw}
\phi(\rho) &\approx&  {\kappa \over e } e^{- i n \theta}  e^{- m |\rho - R|}  \, ,  \nonumber \\
\vec{A} (\rho) &\approx& \hat{\theta} \left[  {\gamma \over 2}
 sgn(\rho - R) \left(1 - e^{-2 m |\rho - R|}  \right)^{\kappa/2 m }
- {n  \over e R}  \right]
\, ,  \end{eqnarray}
with $\gamma = \Phi/ (2 \pi R)$. In first approximation  the energy is dominated by the contribution near the  domain wall; so  according to Eq. (\ref{ew2}) the energy can be approximated by 
\begin{equation}\label{ew3}
E   \approx 2 \pi R \left[ {m \kappa^2 \over e^2}
+  \left({\Phi \over 4 \pi R }  \right)^2 \right] \, . 
\end{equation}

 To obtain a domain wall that is  stable against contraction or expansion
the energy is minimized for a  given magnetic flux. 
Minimizing the  energy as a function of the radius  yields 
\begin{equation}\label{rad}
R \approx { e  \Phi \over 4 \pi \sqrt{ m \kappa} }\, , 
\end{equation}
and thus 
\begin{equation}\label{ew4}
E \approx \sqrt{m \over \kappa } {\kappa^2 \over e }\Phi  \, .
\end{equation}

This value for the energy saturates the  Bogomonl'nyi limit (\ref{ener3})
when $m = \kappa$,  indicating that the fields must be solutions of the 
self-duality equations.  Indeed  we can  verify that  near $r \approx R$  the fields in (\ref{vorw})  solve the  Bogomol'nyi  equations (\ref{eq1})  and (\ref{eq2}) if the  radius $R$  is chosen as in  (\ref{rad}). Using the expression  (\ref{bflux}) for the magnetic flux    we obtain 
$R = ( n + \alpha_n) /(2 \kappa)$ for the radius that minimizes  the energy.
Then,   the condition $R \gg 1/\kappa$ implies $n \gg 1$,  as expected for the  large $n$ limit.

In the domain  wall solution the scalar field reaches its maximum value  $f  =  1 $ at  $\rho = R$. Recalling the discussion
of the previous section, for $ n \gg 1$ the domain wall  approximates the vortex solution in which  the constant  $f_n$  in the short-distance expansion (\ref{sh2})  is chosen as $f_n = f_n^{max}$. 
But for $f_n = f_n^{max}$ the parameter $\alpha_n$ reaches its minimum
value; then according to (\ref{bounds}) we can take 
$\alpha_n = \alpha_n^{min}  \approx n + 1$
and so  the magnetic flux and   the radius become
\begin{eqnarray}\label{phirad}
 \Phi  & \approx &  {4 \pi \over  e } \, \left( n + {1 \over 2}\right)    \, , \nonumber \\   
 R &\approx&  {1 \over \kappa} \left( n + {1 \over 2}\right)  \,  ,
\end{eqnarray}
 respectively. This result  for  $R$ is in   agreement with the one predicted by the maximum of the energy density in Fig. 4.

 The domain wall  solution  can be combined  with the Liouville  approximation   to  find an explicit expression for   the  constant  $f_n^{max}$ .  
The scalar field decays   exponentially away from the domain wall, whereas 
the asymptotic behavior  of  the Liouville solution  shows  a power law both at large and small $\rho$. However,  in the large $n$ limit 
both approaches can be  compatible.   
We introduce the relative coordinate $\xi = R - \rho$,   the domain-wall 
approximation for the scalar field is 
 $ f(r) \approx  \exp{(- \kappa |\xi|)} $. At small $r$ we can use the  power-series solution (\ref{sh2})  or equivalently the Liouville approximation  (\ref{lio}) with $N$ replaced by $n + 1$ and $(\kappa \rho_0)^{(n + 1)}$ by  $2 (n + 1) /f_n$, in either case the leading contribution yields  
\begin{equation}\label{exw}
f(r) \approx f_n^{max}  \kappa^n \left( R - \xi \right)^n = f_n^{max}  \, n^n \, 
\left( 1 + {(1/2)  - \kappa \xi \over n } \right)^n \approx 
f_n^{max} \, n^n \,  e^{[(1/2)- \kappa  |\xi|]} \, . 
\end{equation}
As we are considering the large $n$ limit the last equality 
was  obtained using the  identity 
$\lim_{n \rightarrow \infty}{\left( 1 - {z\over n} \right)^n } = e^{-z} $.
By comparing the  previous result with the domain wall solution  we find that the 
constant in the short-distance expansion (\ref{sh2}) 
of the vortex  solution must  be 

\begin{equation}\label{coex}
f_n^{max} =  { e^{- {1 / 2}} \over n^n} \, .
\end{equation}
This result is expected to provide an adequate  approximation for large $n$.
Remarkably, as   shown in Fig. 5  the actual value of $f_n^{max}$ is 
 reproduced rather well by Eq. (\ref{coex}) for all values of $n$ 
 
The Liouville approximation is also valid at large $\rho$, therefore we can use the 
same approach to determine the coefficient  $C_n$ in the large-distance  expansion (\ref{large}). However,  the large distance expansion  (\ref{large})
and the domain wall solution can only be made compatible if we
take  $\alpha_n   \approx n$ instead  of 
$\alpha_n  = n + 1 $; a sensible approximation for large $n$.  This yields
  $C_n = n^n \approx f^{-1}_n $.  This result 
suggest  the existence of  a  relation between  the large  and short distance 
behavior of the vortex configuration. Indeed,  comparing the leading 
terms in   (\ref{large})  and  (\ref{sh2}) and recalling that 
 $\alpha_n \approx n$,  we find that for large $n$ 
 the vortex configuration 
is symmetric under the exchange
 $ { \kappa \rho \over  n}   \leftrightarrow  {n  \over \kappa \rho}$.

\section{Final remarks}\label{conclu}\indent

In this work we described a self-dual  Maxwell-Chern-Simons model which includes 
an anomalous magnetic   coupling between the scalar and the gauge fields.
We first considered a general scalar potential in order to discuss  both  the symmetric and  the broken  phases  of the theory.  We found that the induced Chern-Simons term  arising from 
the combined effect of the spontaneous symmetry mechanism and the magnetic moment,  has the same properties as those of  the explicit CS term only in the topological trivial sector of the theory.
We  also found that in the  broken phase  the propagating  modes of the gauge field 
consist of two longielliptic waves with different values for the masses.

For a particular relation between the  Chern-Simons  mass and the magnetic moment 
Eq. (\ref{kappag}) the gauge field  equations  reduce from second- to first-order differential equations,
similar to those of the pure Chern-Simons type. The self-dual limit  occurs for a simply $\phi^2$ potential when the scalar and the Chern-Simons masses are equal.  Several properties of the 
self-dual vortices were analyzed both by analytical and  numerical methods.  In particular,  the energy  spectrum of the nontopological  vortices consists  of bands of  constant  width  $\Delta E_n =  2\pi \kappa^2/e^2$ centered at the values $E_n = (4 \pi \kappa^2 /e^2 )\left(n + 3/4 \right)$,  see Eq. (\ref{edesi}).  
  
 We   also considered one-dimensional configurations.  Exact   analytical 
domain wall solutions were found   (Eq. (\ref{sw})) for arbitrary $m$ and $\kappa$. 
In general, these  configurations will not be stationary.
We found, however, that for $m= \kappa$ the  fields in (\ref{sw}) saturate the Bogomol'nyi limit 
 and  consequently the configuration represent a one dimensional stable 
kink with  magnetic flux  and energy per unit length  given by (\ref{exflu}).
Furthermore, we found that in the large flux limit the nontopological vortices
can be correctly approximated  by the domain wall  solution. 

In the self-dual point $m = \kappa$ the vortices become non-interacting and static multisoliton solutions  are expected.   The index-theorem methods can be used to determine the number of independent free parameters that characterize a general $n$-vortex solution of the self-dual equations.  The result can be obtained  by counting the zero modes of the small fluctuations 
which preserve the self-duality equations.  For nontopological vortices the calculation requires 
the subtraction  of the continuous spectrum.   The form of the fluctuations of Eqs. (\ref{sd1}) 
and  (\ref{sd2}) near and far  from the origin are similar to those of solitons in the $\phi^6$ model considered in Ref. \cite{jack2}. The result can then be taken from that paper. The number of free parameters in the general solution of nontopological vortices is $2(n +  \hat{\alpha}_n - 1)$
where $ \hat{\alpha}_n$ is the  greatest integer less that $\alpha_n$. But according to 
(\ref{bounds}) in the $\phi^2$ model  we have  $ \hat{\alpha}_n = n +1 $. Consequently, the number of independent 
free parameters for the $n$-soliton solution is $4 n$.  The  result is consistent with the fact that 
we require $2 n$  parameters to fix the position of $n$ solitons in the plane, 
while the   phases and the fluxes  are determined by the other 
$2 n$ parameters.  

This model raises a number of  interesting questions for further investigation.   
In particular  a complete description of the multisoliton solution  deserve to be clarified. 
It may also be interesting to investigate   the properties 
of the model away from the self-dual point, including the vortex interactions. 
  Finally  will  be of  great interest the study of the properties of the Chern-Simons vortices upon quantization, because they can be considered as candidates for anyon like objects in  planar systems.

\begin{table}
\caption{The parameters  $f_n^{max} $  and  $\alpha_n^{min}$  for the short- and large-expansion of the nontopological vortices as a function  of the vorticity number $n$}\label{tabla}
\begin{tabular} {ccc}
n  &   $ f_n^{max} $  &   $\alpha_n^{min} $  \\
\tableline
1    &   0.607           & 2.142   \\
2    &   0.153          &  3.043   \\
3     &   2.262  $\times$  $10^{-2}$  & 4.021 \\
4     &  2.379  $ \times$  $10^{-3}$  &  5.013  \\
5     &1.946 $\times$  $10^{-4}$   & 6.009    \\
6      & 1.301   $\times$   $10^{-5}$     & 7.009  \\
7     &   7.372  $\times$  $10^{-7}$  & 8.006  \\
8     &   3.623   $\times$  $10^{-8}$  & 9.003  \\
9    &  1.569  $\times$   $10^{-9}$  & 10.002 \\
10  &  6.303  $  \times$  $10^{-11}$     &   11.002 \\
\end{tabular}
\end{table}


\begin{thebibliography}{99}
\bibitem{nota} To  further support the interpretation of $g$ as 
magnetic moment, we    consider the limit in which the scalar  field contribution reduces to a point source charge $i.e.$  $J_{\mu}= eg_{\mu 0}\delta(\vec r\,)$. If we  then combine the usual definition for the magnetic dipole moment with the expression for
${\cal J}_{\mu}$ we obtain 
$ \int {d^2x  \epsilon_{ij}x^i{\cal J}^j} =
-{g\over 2}\epsilon ^{ij}\int {d^2x\epsilon _{jl}x_i\partial ^l
\delta (\vec r)}  
= -g \, .  $
\end{thebibliography}
\end{document}